\documentclass[12pt, draftclsnofoot, onecolumn]{IEEEtran}
\usepackage{graphicx,epstopdf}   
\usepackage{pifont,dsfont}
\usepackage{subcaption}
\usepackage{balance}
\usepackage[utf8x]{inputenc}
\usepackage{color}
\usepackage{epsfig}	
\usepackage{bbm}
\usepackage{xcolor,colortbl}
\usepackage[linesnumbered,algoruled,boxed,lined]{algorithm2e} \usepackage{mathtools,amsthm,amsmath,amsfonts,amssymb}
\usepackage{bm,cite,algorithmic,float}
\usepackage{ulem}  % \sout{}
\usepackage{citesort}

\newtheorem{theorem}{Theorem}

\newtheorem{corollary}{Corollary}

\definecolor{Gray}{gray}{0.9}

\begin{document}
\title{Cooperative Deep Reinforcement Learning for Multiple-Group NB-IoT Networks Optimization
\vspace*{-0.3cm}
}

\author{Nan Jiang, Yansha Deng, Osvaldo Simeone, and Arumugam Nallanathan
\\

\thanks{N. Jiang and A. Nallanathan are with School of Electronic Engineering and Computer Science, Queen Mary University of London, London, UK 
(e-mail:\{nan.jiang, a.nallanathan\}@qmul.ac.uk).}
\thanks{Y. Deng and O. Simeone are with Department of Informatics, King's College London, London, UK (e-mail:\{yansha.deng, osvaldo.simeone\}@kcl.ac.uk).}
}

\maketitle

\vspace*{-0.8cm}

\begin{abstract}
\vspace*{-0.1cm}

NarrowBand-Internet of Things (NB-IoT) is an emerging cellular-based technology that offers a range of flexible configurations for massive IoT radio access from groups of devices with heterogeneous requirements. A configuration specifies the amount of radio resources allocated to each group of devices for random access and for data transmission. Assuming no knowledge of the traffic statistics, the problem is to determine, in an online fashion at each Transmission Time Interval (TTI), the configurations that maximizes the long-term average number of IoT devices that are able to both access and deliver data. Given the complexity of optimal algorithms, a Cooperative Multi-Agent Deep Neural Network based Q-learning (CMA-DQN) approach is developed, whereby each DQN agent independently control a configuration variable for each group. The DQN agents are cooperatively trained in the same environment based on feedback regarding transmission outcomes. CMA-DQN is seen to considerably outperform conventional heuristic approaches based on load estimation.
\color{black}

%\color{olive}{A1: For RACH, the only fairness across different groups is the choosing of repetition value $n^t_{\text{Repe},i}$ that should follows $n^t_{\text{Repe},2}>n^t_{\text{Repe},1}>n^t_{\text{Repe},0}$ [16.3,R1], which has been considered. For the data channel resource scheduling, there is no fairness across different groups. Note that there are a couple of handshakes between eNB and IoT device during the data transmission procedure [Figure 3,R2]. The channel resource will be scheduled for the current handshake immediately whatever the CE group.}
%\\
%\color{olive}{[R1] "Evolved Universal Terrestrial Radio Access (E-UTRA); Physical layer procedures" 3GPP TS 36.213 v14.2.0, 2017}
%\\
%\color{olive}{[R2] Andres-Maldonado, Pilar, et al. "Narrowband IoT Data Transmission Procedures for Massive Machine-Type Communications." IEEE Network 31.6 (2017): 8-15.
%}

\end{abstract}

%========================================================================

\vspace*{-0.0cm}
\section{INTRODUCTION}
\vspace*{-0.0cm}
To effectively support the emerging massive Internet of Things (IoT) ecosystem, the 3GPP has standardized NarrowBand-IoT (NB-IoT), a new radio access technology designed to coexist with Long-Term Evolution \cite{RohdeSchwarz2016white}. NB-IoT supports up to three groups of IoT devices, known as Coverage Enhancement (CE) groups. Each group shares a similar average received Signal-to-Noise Ratio (SNR), as measured based on a broadcast signal, and traffic characteristics (see Fig.1(a)) \cite{wang2017primer}. At the beginning of each uplink Transmission Time Interval (TTI), the evolved Node B (eNB) selects a system configuration that specifies the radio resources allocated to each group in order to accommodate the Random Access CHannel (RACH) procedure with the remaining resources used for data transmission. The key challenge is to optimally balance the allocations of channel resources between the RACH procedure and data transmission so as to provide reliable connections: Allocating too many resources for RACH enhances the random access performance, while leaving insufficient resources for data transmission.  

The eNB observes the number of successful transmissions and collisions on the RACH for all groups at the end of any TTI. This historical information can be potentially used to predict traffic from all groups and to aid the optimization of future TTIs' configurations. Even if one knew all the relevant statistics, tackling this problem in an exact manner would result in a Partially Observable Markov Decision Process (POMDP) with large state and action spaces, which would be generally intractable. The complexity of the problem is compounded by the lack of a prior knowledge at the eNB regarding the traffic and channel statistics.

\vspace*{-0.2cm}
\captionsetup{singlelinecheck=false}  
\begin{figure}[htbp!]
\setlength{\abovecaptionskip}{0pt}
    \begin{center}
        \includegraphics[width=1\textwidth]{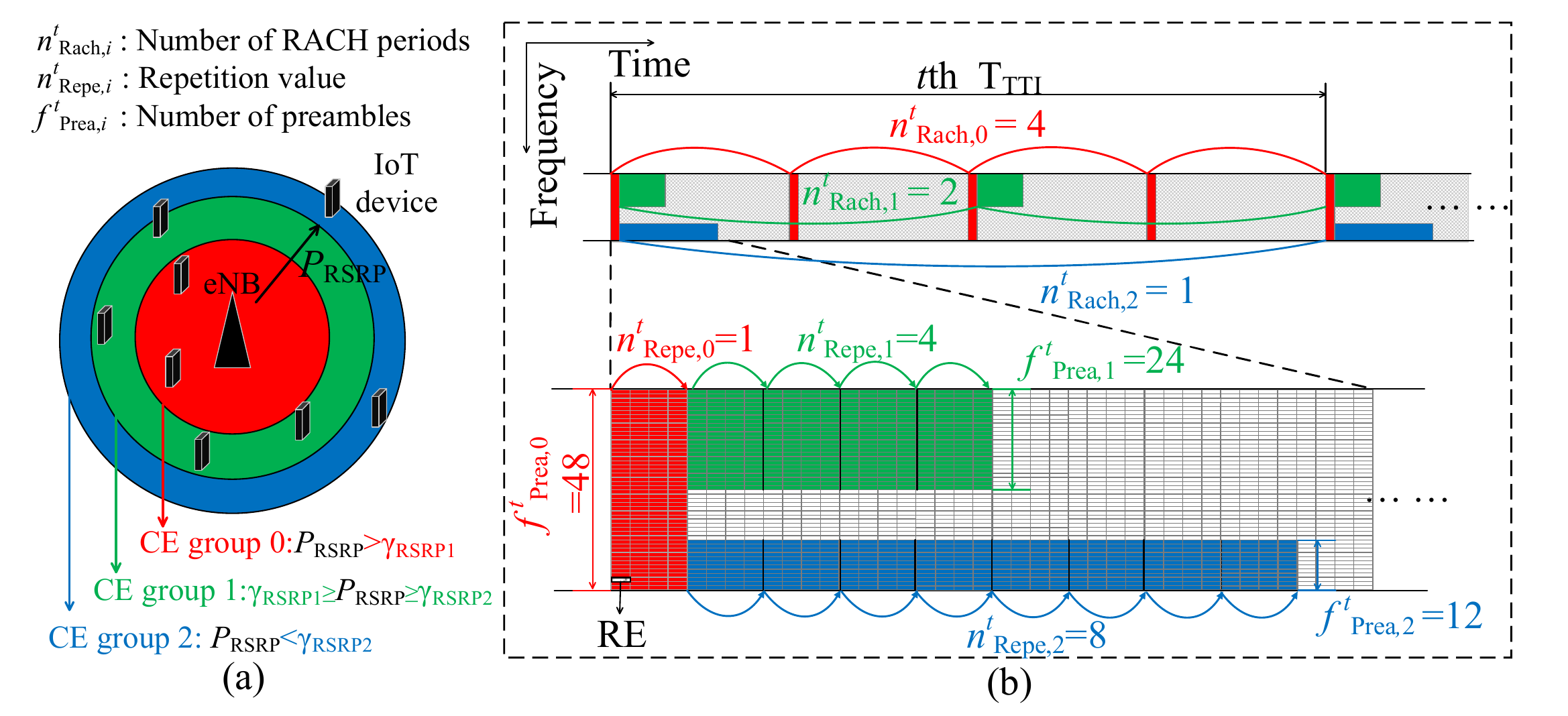}
        \caption{(a) Illustration of system model; (b) Uplink channel frame structure.
        }\label{fig:structure}
    \end{center}
    \vspace*{-0.3cm}
\end{figure}

In light of these challenges, prior works \cite{wiriaatmadja2015hybrid,duan2016d} have tackled the problem under the simplifying assumptions that at most two configurations are allowed and that the optimization is done separately for each group without considering errors due to wireless transmission. In order to consider more complex and practical formulations, Reinforcement Learning (RL) emerges as a natural solution given the availability of feedback in the form of number of successful and unsuccessful transmissions per TTI. Q-learning based Access Class Barring (ACB) schemes have been proposed in \cite{ihun2017A,Luis2018Reinforcement} with the aim of optimizing the access success probability of the RACH. These schemes optimize the ACB procedure by using a tabular approach. Finally, optimizing some of the parameters of the NB-IoT configuration, namely the repetition value (to be defined below), was carried out from the perspective of a single device in terms of latency and power consumption in  \cite{azari2018latency} using a queuing framework. 

%Unfortunately, they set constant reward value that loses flexibility, and the proposed tabular-Q learning framework is incapable in dealing with the high-dimensional variable selection. Besides, whether the Q-learning based approaches \cite{ihun2017A,Luis2018Reinforcement} outperform the conventional resource configuration approaches \cite{duan2016d,wiriaatmadja2015hybrid} is still unknown. 

In this paper, we develop a Cooperative Multi-Agent Deep Neural Network based Q-learning (CMA-DQN) algorithm for online uplink resource configuration in NB-IoT systems. In the proposed approach, a number of DQN agents are cooperatively trained to produce the configurations for the three CE groups. The reliance on Deep Neural Networks (DNNs) addresses the problem of tabular approaches \cite{ihun2017A,Luis2018Reinforcement} in enabling operation over a large state space, while the use of multiple agents deals with the large dimensionality of the output space, corresponding to the configurations for the three CE groups. 

The rest of the paper is organized as follows. Section 2 illustrates the system model. Section 3 discusses the conventional solution. Section 4 presents the CMA-DQN approach. Section 5 provides the numerical results and discussion.

\vspace*{-0.2cm}
\section{SYSTEM MODEL}
\vspace*{-0.1cm}

As illustrated in Fig. \ref{fig:structure}(a), we consider a single-cell NB-IoT network composed of an eNB located at the center of the cell, and a set of static IoT devices randomly located in an area of the plane $\mathbb R^2$. The devices are divided into three CE groups as further discussed below. In each IoT device, uplink data is generated according to random inter-arrival processes over the TTIs, which are Markovian and possibly time-varying as defined in \cite[Ch. 6.1]{3GPP2011Study}.

%The eNB is unaware of the status of these IoT devices of the traffic statistics, and hence it sets the network configuration in each TTI $t$ in terms of RACH and data channel resources according to the predicted upcoming traffic.

\vspace*{-0.45cm}
\subsection{Problem Formulation}
\vspace*{-0.25cm}
Once backlogged, an IoT device executes the contention-based RACH procedure in order to establish a Radio Resource Control (RRC) connection with the eNB. This is done by transmitting a randomly selected preamble for a given number of times within the next RACH period of the current TTI. The RACH process can fail if: (\textit{i}) a collision occurs between two or more IoT devices selecting the same preamble; or (\textit{ii}) there is no collision, but the eNB cannot detect a preamble due to low SNR.
%The RACH procedure for a device may fail due to either of the following two reasons: (\textit{i}) a \textit{collision} occurs when two or more IoT devices select the same preamble; and (\textit{ii}) \textit{SNR outage} occurs when the eNB cannot detect a non-collided preamble due to low SNR. Note that we assume that none of the preambles involved in a collision can be detected following report \cite{3GPP2011Study}, which is different from our previous works about the preamble detection analysis \cite{jiang2018rach,jiang2018collision}. 
\color{blue}
 %(i.e., A collision still occurs in step 3 of RACH when part of the collided preambles cannot be detected following 3GPP report \cite{3GPP2011Study}). Note that this assumption considers all 4 steps RACH, which is diferent from our previous works \cite{jiang2018rach,jiang2018collision} focused on the preamble detection only in RACH step 1.
\color{black}
As shown in Fig. 1(b), for each TTI $t$ and for each CE group $i=0,1,2$, in order to reduce the chance of a collision, the eNB can increase the number $n^t_{\text{Rach},i}$ of RACH periods in the TTI or the number $f^t_{\text{Prea},i}$ of preambles available in each RACH period \cite{3GPP2017PhyCM}. Furthermore, in order to mitigate the SNR outage, the eNB can increase the number $n^t_{\text{Repe},i}$ of times that a preamble transmission is repeated by a device in group $i$ in one RACH period \cite{3GPP2017PhyCM} of the TTI.

%\color{olive} A1: The system configuration is determined in the beginning of each TTI, but this configuration information will be repeatedly broadcast within the whole TTI. Each device might be active in anytime during a TTI. Once a device is active and receives the configuration information, it will transmit the preamble in the next available RACH period immediately. 
%The RACH for a device may fail due to a \textit{collision} occurring when two or more IoT devices select the same preamble, and contention resolution is not taken into account. If no collision occurs, an \textit{SNR} assessment is used to capture the effect of physical channel, where the eNB may not detect a preamble due to low SNR (i.e., we assume a collision still occurs in step 3 of RACH when the collided preambles are not detected following 3GPP report \cite{3GPP2011Study}). Not that this assumption is different from our previous works, which only focus on the preamble detection analysis in step 1 of RACH \cite{jiang2018rach,jiang2018collision}.

After the RRC connection is established, the IoT device requests uplink channel resources from the eNB for control information and data transmission. As shown in Fig. 1(b), given a total number of resources $R_\text{Uplink}$ available for uplink transmission in the TTI, the number of resources available for data transmission is obtained as the difference $R^t_\text{DATA} = R_\text{Uplink}- R^t_\text{RACH}$, where $R^t_\text{RACH}$ is the overall number of Resource Elements (REs)\footnote{The uplink channel consists of 48 sub-carriers within 180 kHz bandwidth. With a 3.75 kHz tone spacing, one RE is composed of one time slot of 2 ms and one sub-carrier of 3.75 kHz \cite{RohdeSchwarz2016white}.} allocated for the RACH procedure. This can be computed as $R^t_\text{RACH} = B_\text{RACH} \sum_{i=0}^{2} n_{\text{Rach},i} n_{\text{Repe},i}  f_{\text{Prea},i}$, where  $B_\text{RACH}$ measures the number of REs required for the transmission of one preamble. 

%\color{red} also, should we specify how many data resources need to be allocated to each devices? 
%\color{olive} A2: The detail about how to calculate the recourse has been described in the end of section 2.2, which has been marked as blue now. Do we need to move this part to here (section 2.1)?

In this work, we tackle the problem of optimizing the RACH configuration defined by parameters $A^t=\{n^t_{\text{Rach},i},f^t_{\text{Prea},i},$ $n^t_{\text{Repe},i}\}_{i=0}^{2}$ for each $i$th group in an online manner for every TTI $t$. In order to make this decision at the beginning of every TTI $t$, the eNB has available for all prior TTIs $t'=1,...,t-1$ the collection $U^{t'}$ consisting of the following variables: the number $V^{t'}_{{\rm cp}}$ of the collided preambles the number $V^{t'}_{{\rm sp}}$ of the successfully received preambles, and the number $V^{t'}_{{\rm ip}}$ of idle preambles  in the $t$th TTI for the RACH, as well as the number $V^{t'}_{{\rm succ}}$ of IoT devices that have successfully sent data and the number $V^{t'}_{{\rm unsc}}$ of IoT devices that are waiting for be allocated data resources. We denote as $O^{t}=\{U^{1},A^{1},U^{2},A^{2},...,U^{t-1},A^{t-1}\}$ the history of all such measurements and past actions. 

The eNB aims at maximizing the long-term average number of devices that successfully transmit data with respect to the stochastic policy $\pi$ that maps the current observation history $O^{t}$ to the probabilities of selecting each possible configuration $A_t$. This problem can be formulated as the optimization
\vspace*{-0.2cm}
\begin{align}\label{q6-1}
&(\text{P1}):  \mathop {\text{max}}\limits_{ \{\pi(A^t|O^t)\}}     \quad   \sum_{k=t}^{\infty} \sum_{i=0}^2  \gamma^{k-t} {\mathbb E}_{\pi} [ V^{k}_{{\rm succ},i} ],
\end{align}
where $\gamma \in [0,1)$ is the discount rate for the performance accrued in future TTIs and index $i$ runs over the CE groups. Since the dynamics of the system is Markovian over the TTI and is defined by the NB-IoT protocol to be further discussed below, this is a POMDP problem that is generally intractable. Approximate solutions will be discussed in Sections 3 and 4.

\vspace*{-0.4cm}
\subsection{NB-IoT Access Network}
\vspace*{-0.15cm}

We now provide additional details on the model and on the NB-IoT protocol. For the wireless channels, we consider the standard power-law path-loss model with path-loss exponent ${\eta}$ and Rayleigh flat-fading. Once an IoT device becomes backlogged, it first determines its associated CE group by comparing the received power of the broadcast signal $P_\text{RSRP}$ to the Reference Signal Received Power (RSRP) thresholds $\{\gamma_\text{RSRP1}, \gamma_\text{RSRP2}\}$ according to the rule \cite{3GPP2017Physical}
\vspace*{-0.1cm}
\begin{align}\label{q3}
 \left\{ \begin{aligned} 
   & \text{CE group 0,} & &\text{if } P_\text{RSRP} > \gamma_\text{RSRP1} ,
 \\& \text{CE group 1,}  & &\text{if } \gamma_\text{RSRP1} \ge P_\text{RSRP} \ge \gamma_\text{RSRP2} ,
 \\& \text{CE group 2,}  & &\text{if } P_\text{RSRP} < \gamma_\text{RSRP2} ,
\end{aligned}  \right. 
\end{align}
where the received power $P_\text{RSRP} = P_\text{NPBCH}  u  ^{ -\eta  }$ is averaged over small-scale Rayleigh fading, $u$ is the device's distance from the eNB, and $P_\text{NPBCH}$ is the broadcast power of eNB  \cite{3GPP2017Physical}. 

After CE group determination, each IoT device in group $i$ repeats a randomly selected preamble $n^t_{\text{Repe},i}$ times in the next RACH period by using a pseudo-random frequency hopping schedule. The preamble consists of four so-called symbol groups, each occupying one RE \cite{lin2016random,jiang2018rach,RohdeSchwarz2016white}. Therefore, a preamble is successfully detected if \textit{at least} one preamble repetition succeeds, which in turn happens if \textit{all} of its four symbol groups are correctly decoded \cite{jiang2018rach}. Assuming that correct detecting is determined by the SNR level $\text{SNR}^t_{\text{sg},j,k}$ for the $j$th repetition and the $k$ symbol group, the correct detecting event $S_{\text{pd}}$ can be expressed as\vspace*{-0.2cm}
\begin{align}\label{q06-4}
S_{\text{pd}}  \buildrel \Delta \over =  \bigcup\limits_{j=1}^{n^t_{\text{Repe},i}} { \Big(  \bigcap\limits_{k=1}^{4}  \big\{ {\text{SNR}^t_{\text{sg},j,k}} \ge \gamma_{\text th} \big\} \Big) },
\end{align}
where $\gamma_{\text th}$ is the SNR threshold, and the SNR can be written as ${\text{SNR}^t_{\text{sg},j,k}} = { P_{\text{RACH},i} {u}^{-\eta } {h}}  /  {\sigma ^2}$ given the preamble transmit power $P_{\text{RACH,}i} =  
  \text{min } \{ {\rho{ u }^{ \eta  }} , \text{ } P_\text{RACHmax} \}  $ for $i=0$ (CE group 0), and $P_{\text{RACH,}i} =  
  P_\text{RACHmax}  $ for $i=1$ or 2. Here, $P_\text{RACHmax}$ is the maximal transmit power of IoT devices. Note that the preamble is transmitted using full path-loss inversion power control for CE group 0 \cite{3GPP2017Physical}, which ensures an average received power of $\rho$ unless the power constraint is violated.

If a RACH fails, the IoT device repeats the procedure until receiving a positive acknowledgement that RRC connection is established, or exceeding $\gamma_{\text{pCE},i}$ RACH attempts while being part of one CE group. If these attempts are exceeded, the device switches to a higher CE group if possible \cite{3GPP2017MAC}. Moreover, the IoT device is allowed to attempt the RACH procedure no more than $\gamma_{\text{pMax}}$ times before dropping a packet.

%\color{olive} A2: The IoT device will not receive a negative acknowledgement from eNB. This is due to that the eNB may never received any information from this device due to the low received SNR. In this scenario, the IoT device determines not to retry to access to this eNB by itself, and it will start an eNB selection procedure that tries to search other eNBs (i.e., we do not consider this here). Another is I am worried about whether "time-out period" is accurate or not here, due to that IoT device only counts the times of RACH attempts rather than a period. This is due to that the eNB may execute a back-off scheme (i.e., not consider in our work), so the time-out period can be quite changeable due to different back-off factor.
%\color{black}

To allocate data resources among the devices that have correctly completed the RACH procedure, we adopt a basic random scheduling strategy, whereby an ordered list of all devices that have successfully completed the RACH procedure but have not received a data channel is compiled using a random order. In each TTI, devices in the list are considered in order for access to the data channel until the data resources are insufficient to serve the next device in the list. The remaining RRC connections between the unscheduled IoT devices and the eNB will be preserved within at most $\gamma_\text{RRC}$ subsequent TTIs, and attempts will be made to schedule the device's data during these TTIs \cite{3GPP2017Requirements}. The condition that the data resources are sufficient in TTI $t$ is expressed as
\vspace*{-0.4cm}
\begin{align}\label{q5-1}
{R}^t_{\text{DATA}} \ge \sum_{i=0}^{2} {r}^t_{\text{DATA},i} V^t_{\text{sch},i},
\end{align}
where $\sum_{i=0}^{2} V^t_{\text{sch},i}\leq \sum_{i=0}^{2} (V^t_{\text{sp},i} + V^{t-1}_{\text{unsc},i})$ is the number of scheduled devices; ${r}^t_{\text{DATA},i} = B_\text{DATA}  \times  n^t_{\text{Repe},i}$ is the number of required REs for serving one IoT device within the $i$th CE group; and $B_\text{DATA}$ is the number of REs per repetition. Note that the number of repetitions is the same as for preamble transmission \cite{RohdeSchwarz2016white}.

\vspace*{-0.2cm}
\section{Conventional Solutions}
\vspace*{-0.1cm}

Due to its complexity, most previous works simplify the optimization in (\ref{q6-1}) by considering the greedy formulation 
%\vspace*{-0.2cm}
\begin{align}\label{q6-2}
&(\text{P2}):  \quad \mathop {\text{max}}\limits_{ \pi (A^t|O^t )}   \quad  {\mathbb E}_{\pi} [ V^{t}_{{\rm succ},i} ],
\end{align}
for some group $i$, whereby only the performance in the current TTI is considered. Furthermore, the expectation in (5) is approximated based on an estimate of the load in TTI $t$ as discussed below; and the action space for $A_t$ is typically reduced to include only some parameters such as the number $f^t_{\text{Prea},i}$ of preambles in each RACH period \cite{duan2016d,wiriaatmadja2015hybrid}. 

To elaborate, we now briefly describe a solution based on \cite{duan2016d} that follows the outlined simplifying principles. We drop the group index $i$ in order to avoid unnecessary notation. At the beginning of each TTI, the scheme first estimates the number ${\hat D}^t_{\text{RACH}}$ of IoT devices that will attempt RACH access in the $t$th TTI, and then adjusts only the parameters $A_t=f^t_{\text{Prea}}$ according to the estimated load. The estimate is given as
\vspace*{-0.2cm}
\begin{align}\label{1q10}
& {\hat{D}^t_{\text{RACH}}}  =  \text{max} \big\{  2V^{t-1}_{{\rm coll}},  \zeta^{t-1} + \delta^{t} \big\},
\end{align}
where the term $2V^{t-1}_{{\rm coll}}$ reflects the fact that there are at least $2V^{t-1}_{{\rm coll}}$ IoT devices colliding in the last TTI; $\delta^{t} \approx \delta^{t-1} = \hat{D}^{t-1}_{\text{RACH}} − \hat{D}^{t-2}_{\text{RACH}}$ is the difference between the estimated numbers of RACH attempting IoT devices in the ($t-1$)th and the $t$th TTIs \cite{duan2016d}; and $\zeta^{t-1} = \text{log}_{ ({{f^{t-1}_{\text{Prea}}}-1} ) / {f^{t-1}_{\text{Prea}}} }({V^{t-1}_{{\rm idle}}} / {f^{t-1}_{\text{Prea}}})$ is an estimate of the number of RACH-attempting IoT devices in the ($t-1$)th TTI obtained via moment matching \cite{duan2016d}. 

%$\zeta^{t-1} = \text{log}_{\frac{{f^{t-1}_{\text{Prea}}}-1}{f^{t-1}_{\text{Prea}}}}(\frac{V^{t-1}_{{\rm idle}}}{f^{t-1}_{\text{Prea}}})$
%\color{olive} A3: This formula is obtained by calculating the inverse of the function of expected preambles experiencing idles ${\mathbb E} \{ {\cal V}^{t-1}_{{\rm idle},0} = {f^{t-1}_{\text{Prea},0}}  \big(1-\frac{1}{f^{t-1}_{\text{Prea},0}} \big)^{n} $. I omitted it due to the page limitation. Do you think we need to give more clear explanation of it?
%\color{black}

Using the estimated load given in (\ref{1q10}), the approach, which is referred to as Load Estimation based Uplink Resource Configuration (LE-URC), attempts to solve problem (\ref{q6-2}) by approximating the objective as  
\vspace*{-0.2cm}
\begin{align}\label{1q5}
 {\mathbb E}_{\pi} [ V^{t}_{{\rm succ}}] & \approx  \text{min}\{{\mathbb E} \{ {V}^{t}_{{\rm reqs}} \big| {\hat{D}^t_{\text{RACH}}}  \} , { V}^{t}_{up}\},
\end{align}
where ${\mathbb E} \{ V^{t}_{{\rm reqs}} \big| {\hat{D}^t_{\text{RACH}}}= n  \} = n \big(1-\frac{1}{f^t_{\text{Prea}}} \big)^{n-1} + V^{t-1}_{\text{unsc}} $ is the expected number of IoT devices requesting uplink resource in the $t$th TTI; and ${V}^{t}_{\text{up}} = \frac{R_{\rm Uplink} - R^t_{\rm RACH}}{{r}^t_{\text{DATA}}} $ is an upper bound on the number of IoT devices can be scheduled.

\vspace*{-0.1cm}
\section{Cooperative Multi-agent DNN-Q Approach}
\vspace*{-0.1cm}
We now introduce an RL-based approach to tackle problem (\ref{q6-1}). A direct application of the DQN approach \cite{mnih2015human} or of its enhancement proposed in \cite{van2016deep}, whereby the policy $\pi(A^t|O^t)$ for all $t$ is modelled by a DNN, is not feasible due to the increasing size of the action $A^t$. In order to overcome this issue, we break up the action space by considering separately each of the nine action variables in $A_t$. Recall that we have three variables for each group $i$, namely ${n}_{\text{Rach},i}$, ${n}_{\text{Repe},i}$, and ${f}_{\text{Prea},i}$.

A separate DQN agent is introduced for each output variable in $A^t$. We define as $A^t_k$ the action selected by the $k$th agent. Each $k$th agent is responsible to update the value $Q(S^t,A_k^t)$ of action $A_k^t$ in state $S^t$, where the state variable $S^t=[U^{t-M_o},A^{t-M_o},...,U^{t-1},A^{t-1}]$ only includes information about the last $M_o$ TTIs. All agents receive the same reward signal $R^t=\sum_{i=0}^{i=2} V^t_{\text{succ},i}$ at the end of each TTI as per problem (\ref{q6-1}). The use of this common reward signal ensures that all DQN agents aim at cooperatively increase the objective in (\ref{q6-1}). Note that the approach can be interpreted as applying a factorization of the overall value function akin to the approach proposed in \cite{Sunehag2018} for multi-agent systems.

The DQN agents are trained in parallel. Each agent $k$ parameterizes the action-state value function $Q(S^t,A_k^t)$ by using a function $Q(S, A_k;\theta_k)$, where $\bm{\theta}_k$ represents the weights matrix of a DNN with fully-connected layers. The input of the DNN is given by the variables in state $S^t$; the intermediate layers are Rectifier Linear Units (ReLUs); while the output layer is composed of linear units. Each output neurons provides the value of one of the actions in $A^t_k$ as in \cite{mnih2015human}. The weights matrix $\theta_k$ is updated online along each training episode by using double deep Q-learning (DDQN) \cite{van2016deep}. Accordingly, learning takes place over multiple training episodes, with each episode of duration $N_\text{TTI}$ TTI periods. In each TTI, the parameters $\theta_k$ of the Q-function approximator $Q(S^t,A^t_k;\theta_k)$ are updated using Stochastic Gradient Descent at all agents $k$ as
\vspace*{-0.2cm}
\begin{align}\label{q19-2}
\theta_k^{t+1} =  \theta_k^{t}  -   \alpha \nabla L_k(\theta^t_k),
\end{align}
where $\alpha$ is RMSProp learning rate \cite{tieleman2012lecture}, $\nabla L_k(\theta_k)$ is the gradient of the loss function $L_k(\theta^t_k)$ used to train the Q-function approximator. This is given as
\vspace*{-0.15cm}
\begin{align}\label{q19}
 \nabla L_k(\theta_k^t) =  &  {\mathbb E}_{S^i,A_k^i,R^{i+1},S^{i+1}} \big[\big( R^{i+1} + \gamma  \mathop{\text{max}}\limits_{{A_k}}Q(S^{i+1}, 
 \nonumber \\ 
 & \hspace{-0.5cm} A_k; \bar{\theta}^t_{k}) -  Q(S^i, A_k^i;   \theta^t_k) \big) \nabla_{\theta_k} Q(S^i, A^i_k; \theta^t_k)\big],
\end{align} where the expectation is taken with respect to randomly selected previous samples $(S^i,A^i_k,S^{i+1},R^{i+1})$ for some $i\in\{t-M_r,...,t+1\}$, with $M_r$ being the replay memory \cite{mnih2015human}. When $t-M_r$ is negative, this is to be intended as including samples from the previous episode. Following DDQN \cite{van2016deep}, in (\ref{q19}), $\bar{\theta}^t_{k}$ is a so-called target parameter that is used to estimate the future value of the Q-function in the update rule. This parameter is periodically copied from the current value $\theta^t_{k}$ and kept fixed for a number of episodes.

\vspace*{-0.4cm}
\section{Simulation Results and Discussion}
\vspace*{-0.2cm}

In this section, we evaluate the performance of the proposed CMA-DQN and compare it with the conventional LE-URC described in Sec. 3 via numerical experiments. The eNB is assumed to be located at the center of a circle area with 12 km radius, and we adopt the standard network parameters listed in Table \ref{table_2} following \cite{RohdeSchwarz2016white,wang2017primer,3GPP2015Cellular,3GPP2017MAC,3GPP2017PhyCM}. Accordingly, one epoch consists of 937 TTIs (i.e., 10 minutes). Throughout epoch, each device have a bursty traffic profile, where the packet generation probability is given by the time limited Beta profile defined in \cite[Ch. 6.1]{3GPP2011Study} with parameters $(3,4)$, which has a peak around the 400th TTI. The resulting average number of generated packets is shown as dashed line in Fig. \ref{fig:2}. The DQNs used by CMA-DQN have three hidden layers, each with 128 ReLU units, where other hyperparameters are listed in Table \ref{table_2}. All results are obtained by averaging over 1000 training episodes. 
\captionsetup{singlelinecheck=true}
\begin{table}[htbp!]
	\centering
	\caption{Simulation Parameters and Q-learning hyperparameters
	\vspace*{-0.3cm}}
	{\renewcommand{\arraystretch}{0.6}
		\begin{tabular}{|*{1}{p{4.8cm}}|*{1}{p{2.6cm}} |*{1}{p{4.8cm}}|*{1}{p{2.6cm}|} }
			\hline
			\rowcolor{Gray}
		   \bf{Simulation Parameters}   &    \bf{Setting}$\vphantom{\Big(}$ &  \bf{Simulation Parameters}   &    \bf{Setting}$\vphantom{\Big(}$ \\  \hline 
            Path-loss exponent $\eta$ $\vphantom{\big(}$ & 4 
            & Noise power $\sigma^2  $ $\vphantom{\Big(}$ &   -138 dBm  \\
               $\vphantom{\big(}$Received SNR threshold  $\gamma_{\rm th}$  & 0 dB
            &   Power control threshold $\rho$ $\vphantom{\big(}$  & 120 dB  \\
            $\vphantom{\big(}$eNB broadcast power $P_{\text{NPBCH}}$&  35 dBm 
            &   TTI$\vphantom{\big(}$& 640 ms  \\
            $\vphantom{\big(}$Bursty traffic duration & 10 mins 
		    & $\vphantom{\big(}$IoT devices & 30000 \\
           Maximum transmit power $P_{\text{RACHmax}}\vphantom{\big(}$ &  23 dBm  
		    & Set of number of preambles ${\cal F}_\text{Prea}$ $\vphantom{\big(}$ & \{12,24,36,48\}   \\
		  $\vphantom{\big(}$Maximum resource requests $\gamma_\text{RRC}$ & 5 
			&  Set of repetition value ${\cal N}_\text{Repe}$ $\vphantom{\big(}$ &  \{1,2,4,8,16,32\} \\ 
	      $\vphantom{\big(}$Maximum RACH in one CE $\gamma_{\text{pCE},i}$ & 5
            & Set of RACH periods ${\cal N}_\text{Rach}$ $\vphantom{\big(}$ & \{1,2,4\}  \\
            	               $\vphantom{\big(}$Maximum RACH attempts $\gamma_{\text{pMax}}$  & 10
           & RSRP threshold \{$\gamma_\text{RSRP1}$,$\gamma_\text{RSRP2}$\}& \{0,-5\} dB \\ 
            $\vphantom{\big(}$REs required for $B_\text{RACH}$ & 4
           & REs required for $B_\text{DATA}$ & 32  \\ 
			\hline
			\rowcolor{Gray}
		   \bf{Q-learning Hyperparameters}   &    \bf{Value}$\vphantom{\Big(}$ &  \bf{Q-learning Hyperparameters}   &    \bf{Value}$\vphantom{\Big(}$ \\  \hline
		   $\vphantom{\big(}$  Exploration $\epsilon$  & [0.1,1]
		   & RMSProp Learning rate $\lambda_\text{RMS}$ & 0.0001 \\
		   $\vphantom{\big(}$Discount rate $\gamma$ & 0.5 
           & Minibatch size $\vphantom{\big(}$ & 32 \\
           $\vphantom{\big(}$Replay memory & 10000
           & Target Q-network update frequency & 1000 \\
                        \hline
		\end{tabular}
	}
	\vspace*{-0.1cm}
	\label{table_2}
\end{table}

Fig. \ref{fig:2} compares the number of successfully served IoT devices $V_\text{succ}$ during one epoch using CMA-DQN and LE-URC. The ``LE-URC-[1,4,8]" and ``LE-URC-[2,8,16]" curves represent the LE-URC approach with the repetition values $\{n_{\text{Repe},0},n_{\text{Repe},1},n_{\text{Repe},2}\}$ set to $\{1,4,8\}$ and $\{2,8,16\}$, respectively. We observe that the CMA-DQN slightly outperforms LE-URC in the light traffic regions at the beginning and end of the epoch, but it substantially outperforms LE-URC in the period of heavy traffic in the middle of the epoch. This demonstrates the capability of CMA-DQN to better manage the scarce channel resources in the presence of heavy traffic. It is also observed that increasing the repetition value of each CE group with LE-URC improves the received SNR, and thus the RACH success rate, in the light traffic region, but it degrades the scheduling success rate due to limited channel resource in the heavy traffic region.

\captionsetup{singlelinecheck=false} 
\begin{figure}[htbp!]
\setcounter{figure}{1} 
\vspace*{-0.5cm}
    \begin{center}
        \begin{minipage}[t]{0.9\textwidth}
    \centering
        \includegraphics[width=1\textwidth]{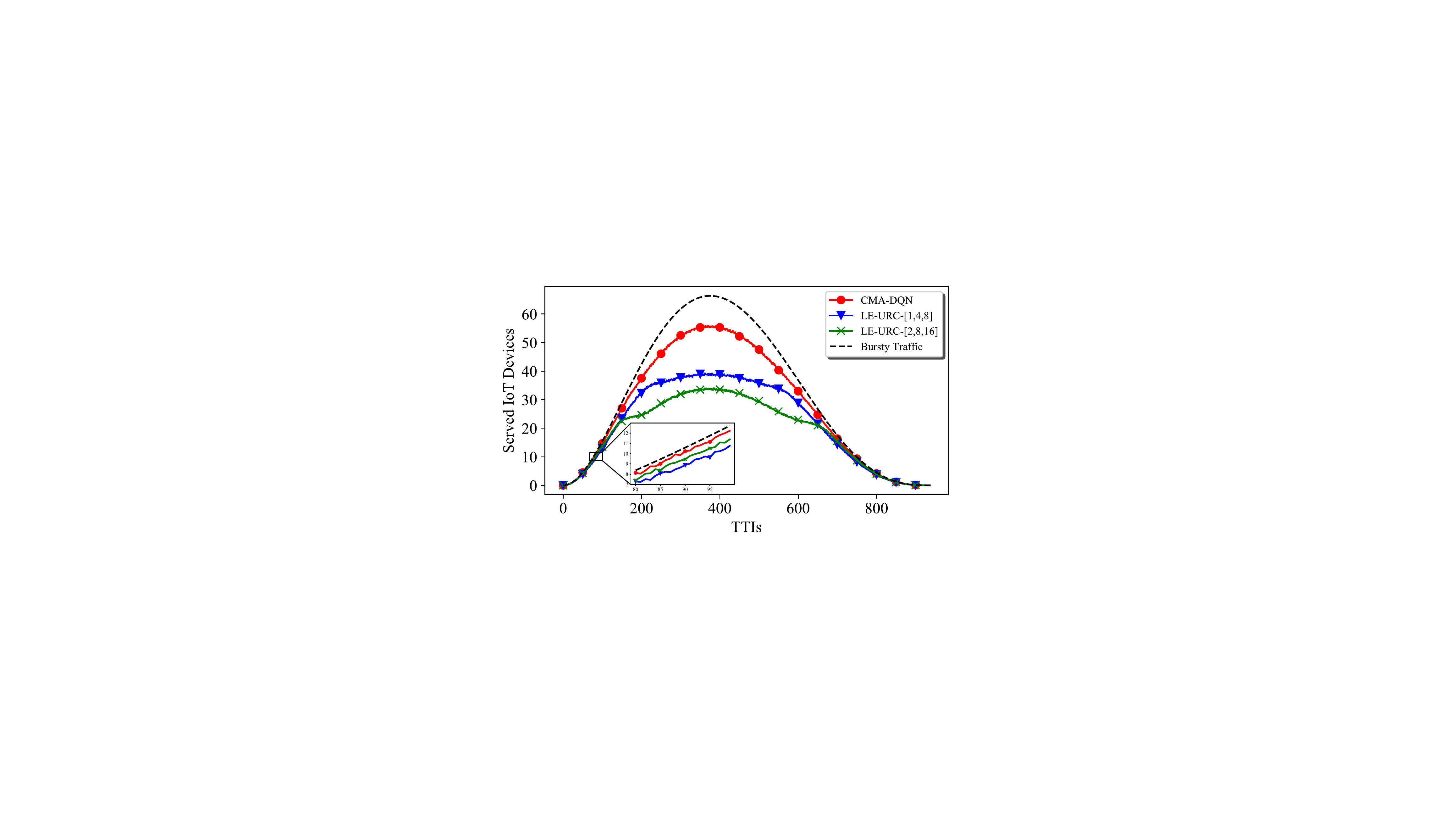}
        \vspace*{-0.9cm}
        \caption{\scriptsize The average number of successfully served IoT devices $V_\text{succ}$ per TTI during one bursty traffic duration. The dashed line represents the average number of generated packets per TTI.}
                \label{fig:2}
        \end{minipage}
    \end{center}
    \vspace*{-0.5cm}
\end{figure}

To gain more insight into the operation of CMA-DQN, Fig. \ref{fig:3} plots the average number $n^t_{\text{Repe},i}$ of repetitions and the average number of Random Access Opportunities (RAOs), defined as the product $n^t_{\text{Rach},i} \times f^t_{\text{Prea},i}$, for each CE group $i$ that are selected by CMA-DQN over the training episodes. As seen in Fig. \ref{fig:3}(a)-(c), CMA-DQN increases the number of repetitions in the light traffic region in order to improve the SNR and reduce RACH failures, while decreasing it in the heavy traffic region so as to reduce scheduling failures. As illustrated in Fig. \ref{fig:3}(d)-(f), this allows CMA-DQN to increase the number of RAOs in the high traffic regime mitigating the impact of collisions on the throughput. In contrast, for the CE groups 1 and 2, in the heavy traffic region,
 LE-URC decreases the number of RAOs in order to reduce resource scheduling failures, causing an overall lower throughput as seen in Fig. \ref{fig:2}.

\vspace*{-0.0cm}
\captionsetup{singlelinecheck=false} 
\begin{figure}[htbp!]
\setcounter{figure}{2} 
    \begin{center}
        \begin{minipage}[t]{1\textwidth}
    \centering
        \includegraphics[width=1\textwidth]{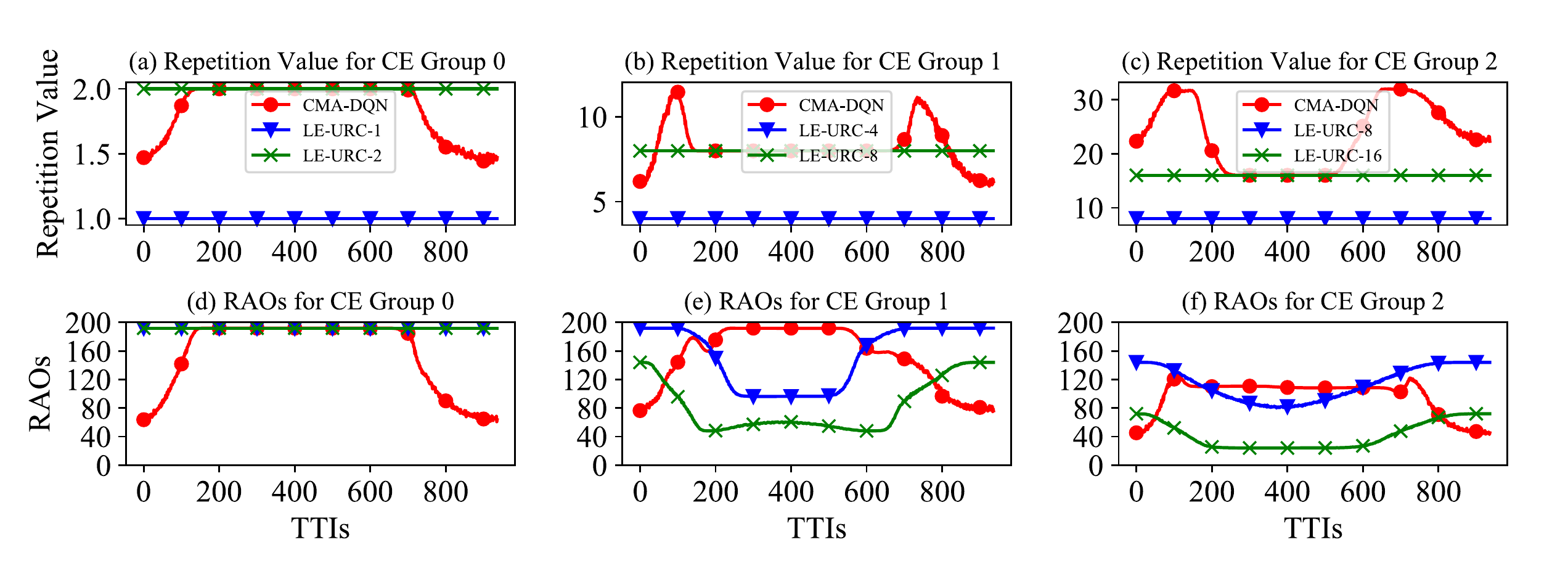}
        \vspace*{-0.9cm}
        \caption{\scriptsize The allocated repetition value $n^t_{\text{Repe},i}$, and RAOs producted by $n^t_{\text{Rach},i} \times f^t_{\text{Prea},i}$.}
                \label{fig:3}
        \end{minipage}
    \end{center}
    \vspace*{-1.0cm}
\end{figure}

%to reduce scheduling failures in the heavy traffic region, each approach decreases the number of allocated preambles of CE group 2 after a traffic threshold, and even LE-URC approaches decreases that of CE group 1. Furthermore, the allocating preambles of CMA-DQN approach experiences the mildest decreasing trend, due to that this approach intelligently well-serves massive IoT devices with extended coverage over the whole bursty traffic duration that leads to relatively light traffic burden in each TTI.

% Overall, the proposed CMA-DQN approach can dynamically optimize the number of served IoT devices in the high-complexity scenarios, which intelligently well-serve massive IoT devices with extended coverage in NB-IoT networks.

%\section{CONCLUSION}

%In this paper, we developed a DNN-Q algorithm via centralized multi-agent cooperation approach, which can optimize the number of served IoT devices in real-time in NB-IoT networks. The network configurations are independently determined by different DNN-Q agents training in the same environment with the shared decision status, thereby solves the problem that DRL algorithm do not converge in high-dimensional configurations. Our results demonstrated that the proposed CA-DNN-Q approaches considerably outperform the conventional LE-URC approach in terms of the number of served IoT devices.

\bibliographystyle{IEEEtran}
\bibliography{IEEEabrv,RA_bib}

\end{document}